\documentclass[aps,prl,superscriptaddress,twocolumn,amsmath,amssymb]{revtex4}
\usepackage{graphicx} 
\usepackage{graphics}
\usepackage{bm}       
\usepackage{epsfig}   
\newcommand{\ket}[1]{|{#1}\rangle}
\newcommand{\bra}[1]{\langle{#1}|}
\begin{document}


\title{Experimental demonstration of four-party quantum secret sharing}

\author{S. Gaertner}
\affiliation{Sektion Physik, 
             Ludwig-Maximilians-Universit\"at,
             80799 M\"unchen, Germany}
\affiliation{Max-Planck-Institut f\"ur Quantenoptik, 
             D-85748 Garching, Germany}
\author{C. Kurtsiefer}
\affiliation{Department of Physics, 
             National University of Singapore, 
             117542 Singapore, Singapore}
\author{M. Bourennane}
\affiliation{Physics Department, 
             Stockholm University, 
             S-10691 Stockholm, Sweden} 
\author{H. Weinfurter}
\affiliation{Sektion Physik, 
             Ludwig-Maximilians-Universit\"at,
             80799 M\"unchen, Germany}
\affiliation{Max-Planck-Institut f\"ur Quantenoptik, 
             D-85748 Garching, Germany}


\begin{abstract}
Secret sharing is a multiparty cryptographic task in which some secret 
information is splitted into several pieces which are distributed among the 
participants such that only an authorized set of participants can reconstruct 
the original secret. Similar to quantum key distribution, in quantum secret 
sharing, the secrecy of the shared information relies not on computational 
assumptions, but on laws of quantum physics. Here, we present an experimental 
demonstration of four-party quantum secret sharing via the resource of 
four-photon entanglement. 
\end{abstract}
\maketitle
Entanglement is a fundamental resource for many quantum communication schemes.
Bipartite entanglement has been used for the experimental demonstration of 
two-party quantum communication schemes like quantum dense coding~\cite{qdc}, 
quantum teleportation~\cite{qtp} or entanglement based quantum 
cryptography~\cite{ebqc}. Similar to such two-party quantum communication 
schemes, multipartite entanglement allows the experimental implementation of 
multiparty quantum communication schemes, like multiparty quantum 
teleportation~\cite{mpqtp}, quantum telecloning~\cite{qtc}, multiparty quantum
key distribution or quantum secret sharing (QSS)~\cite{qss_ci_01}. 

Secret sharing was introduced independently in $1979$ by Shamir and Blakely
~\cite{ss_01,ss_02}. In a secret sharing scheme between $n$ participants a 
designated party, usually called the dealer, splits some secret information 
into $n-1$ shares and distributes these shares to each participant in such a 
way that only a particular authorized set of participants can reconstruct that
secret information. The qualified subsets of participants who can recover the 
secret are called access sets, subsets which have no information about the 
secret are called non-access sets and subsets which have partial information 
are called semi-access sets~\cite{npss}. 
Secret sharing has many different applications, e.g. the management of 
cryptographic keys, the establishment of access codes with restricted access 
and as a component of secure multiparty computation.

In contrast to classical secret sharing, the information splitting of a secret
and the information distribution in QSS is realized by local measurements on 
distributed quantum states. Therefore, QSS allows to distribute the shares 
securely in the presence of eavesdropping. Moreover, QSS distinguishes further
between QSS of classical information~\cite{qss_ci_01} and QSS of quantum 
information~\cite{qss_qi}. Here, we consider QSS of classical information. 
Different protocols for three and four-party QSS using the resource of 
multipartite entanglement have been proposed~\cite{qss_ci_01,qss_ci_03}. But, 
until now, only three-photon entanglement has been used to proof the 
experimental feasibility and to give an experimental demonstration of 
three-party QSS~\cite{qss_exp_01,qss_exp_02}.

In this Letter, we present the first experimental demonstration of four-party 
QSS via four-photon entanglement. In this scheme, any one of the four 
participants can act as the dealer, while the remaining three participants 
form the access set. For the experimental implementation, we use the following 
four-photon polarization-entangled state~\cite{psi4,psi4m}:
\begin{eqnarray}
\label{psi4-}
\ket{\Psi^{-}_{4}}
&=&\frac{1}{2\sqrt{3}}
   \left[\rule{0em}{1.0em}\right. 2\ket{HHVV}-\ket{HVHV}-\ket{HVVH}\\
& &\quad\quad-\ket{VHHV}-\ket{VHVH}+2\ket{VVHH}\left. 
\rule{0em}{1.0em}\right]_{abcd} \nonumber
\end{eqnarray}
where $H$ and $V$ denotes horizontal and vertical polarization of photons 
in the four spatial modes $a,b,c$ and $d$. This state shows perfect 
four-photon polarization correlations as indicated by the four-photon 
polarization correlation function defined as the expectation value of the 
product of the four operators
$\hat{s}_{x}=\ket{+,\Phi}_{x}\bra{+,\Phi}-\ket{-,\Phi}_{x}\bra{-,\Phi}\,$, 
with eigenstates 
$\ket{\pm,\Phi}_{x}=1/\sqrt{2}(\ket{R}\pm e^{i\phi_{x}}\ket{L})$ 
and eigenvalues $\pm 1$, where $R$ and $L$ denote right- and lefthanded 
circular polarization. The explicit expression of this correlation function 
for the four-photon state given by Eq.~(\ref{psi4-}), is: 
\begin{eqnarray}
\label{cor4f}
E(\phi_a,\phi_b,\phi_c,\phi_d)&=&{2\over3}\cos(\phi_a+\phi_b-\phi_c-\phi_d)\\
&&+{1\over3}\cos(\phi_a-\phi_b)\cos(\phi_c-\phi_d)\,.\nonumber
\end{eqnarray}
Another property of this state, which is useful for the realization of 
multiparty secure quantum communication, is its ability to violate a 
four-party Bell inequality~\cite{mzbi}:
\begin{eqnarray}
\label{mzbi}
S={1\over2^4}\sum\limits_{s_{x}=\pm 1}
\left|
\sum\limits_{y=1,2} s^{k}_{a} s^{l}_{b} s^{m}_{c} s^{n}_{d}\, 
E(\phi_{a}^k,\phi_{b}^l,\phi_{c}^m, \phi_{d}^n) 
\right| \le 1 
\end{eqnarray}
where each index $y=k,l,m,n$ denotes a pair of angles defining the local 
polarization measurement settings required for the evaluation. A strong 
violation of this Bell inequality (with $S=1.886$) is obtained e.g. for  
$\phi^{1,2}_{b}=0,\pi/2$ and $\phi^{1,2}_{a,c,d}=\pi/4,-\pi/4$, or any other
setting resulting by permutation of the indices.

The four-party QSS protocol works as follows: 
Alice, Bob, Claire and David share each a photon from the $\ket{\Psi_{4}^{-}}$
state. In the following, we assume that Alice is the dealer. Each party chooses
randomly between two complementary measurement bases. This can be for example 
either the $\{H,V\}$ and $\{P,M\}$ basis 
($\phi_{x}=0$, $\pi/2$) analog to the BB84 protocol~\cite{bb84}, or a basis 
set like $\{\{22.5^{\circ},112.5^{\circ}\},\{-22.5^{\circ},67.5^{\circ}\}\}$ 
$(\phi_{x}=\pi/4,-\pi/4)$, which can also be used to violate the Bell 
inequality given by Eq.~(\ref{mzbi}), in analogy to the Ekert scheme
~\cite{e91}. To transfer the measurement results into a key sequence, each 
participant identifies his result either with a bit value of $0$ or $1$. The 
measurements will be repeated until they have established a raw key of desired
length. For key sifting, each participant announces publicly whenever he has 
registered a photon and which measurement basis he has used, but not the 
results. For the announcement, Bob, Claire and David send their information 
about their measurement to Alice. After that, Alice can decide which quadruples
will be used for secure communication: According to Eq.~(\ref{cor4f}), she will
keep those results where all participants used the same basis setting and will
drop all others.

The information splitting works as follows: Consider e.g the case where all 
four parties have measured in the $\{H,V\}$ basis. Suppose Bob obtained the 
result $\ket{H}_{b}$ $(\ket{V}_{b})$, then he can not predict with certainty 
the measurement result of Alice, because there are two different possible 
outcomes for her. Similarly, each other participant is not able to obtain the 
secret of Alice without cooperation. Let us now consider pairwise cooperation.
If Bob and Claire obtained the result $\ket{HV}_{bc}$ $(\ket{VH}_{bc})$, both 
together are not able to get the secret. They need the help of David to 
determine the key bit of Alice. Only in cases where Bob and Claire have 
obtained the result $\ket{HH}_{bc}$ $(\ket{VV}_{bc})$ they can infer the key 
bit without the help of David. The same is true for Bob and David, or Claire 
and David. This implies that always two of them have some partial information 
on Alice's key. However, Bob, Claire and David must cooperate to retrieve 
the complete key and to ensure perfect information-theoretic security, the 
partial information of any semi-access set can be removed by the application 
of a hash function (privacy amplification)~\cite{pa,krr}. 

After key sifting, all participants have to check for external eavesdropping. 
Depending on their chosen basis settings they can proceed as follows: If they 
used a basis set similar to the BB84 protocol, they can use a fraction of 
their measurement results (which should be perfectly correlated) for the 
evaluation of the quantum bit error rate (QBER) defined as the ratio of wrong 
bits to all bits. If the QBER is low enough, they can use their results to 
distill a secure key, otherwise they have to discard their bits~\cite{tv4qss}. 
If they used a basis set which can be used to violate the Bell inequality 
given by Eq.~(\ref{mzbi}), they can use a fraction of their measurement 
results for the evaluation. If the violation is high enough, they can use 
their key bits, otherwise they have to drop them~\cite{bv4qss}.

To finally obtain a common secure key, they have to perform key reconciliation 
and privacy amplification. For this, different strategies developed for 
quantum key distribution (QKD) can be adapted~\cite{kr,pa}. After that, Alice 
can use the final key for an unbreakable encryption of her secret information 
via the Vernam cipher~\cite{vc} and broadcasts the encrypted message to Bob, 
Claire and David. Cooperation of Bob, Claire and David allows the 
reconstruction of the key and therefore to obtain the secret information of 
Alice.

\begin{figure}[h!]
\begin{center}
\epsfig{file=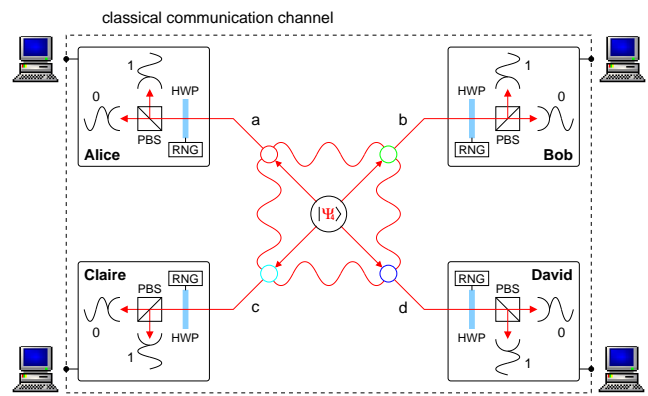, width=8.5cm}
\caption{\label{setup}
Scheme of the experimental implementation: 
The four-photon source distributes the four photons of the entangled state 
$\ket{\Psi^{-}_{4}}$ into the four modes $a,b,c$ and $d$. Each party uses a 
random number generator (RNG) which sets a half-wave plate (HWP) angle and a 
polarizing beam splitter (PBS) followed by two avalanche photodiodes for the 
analysis and registration of the distributed photons.}
\end{center}
\end{figure}

The four-party QSS protocol was experimentally implemented as sketched in 
Fig.~\ref{setup}. We used a mode-locked Ti:Sapphire laser emitting pulses with
a pulse length of about $120$~fs at a repetition rate of $82$~MHz. This 
radiation is frequency-doubled with a lithium-triborate crystal to 
$\lambda=390$~nm which is used to pump a $2$~mm thick beta-barium-borate 
crystal to generate the four-photon polarization-entangled state given by 
Eq.~(\ref{psi4-}) via pulsed type-II parametric down-conversion.
This state results directly from the four-photon emission of the pulsed 
down-conversion source and the usage of two beam splitters to distribute these
photons into the four spatial modes $a,b,c$ and $d$~\cite{psi4m,psi4p}.
Alice, Bob, Claire and David obtain each a photon from the four-photon state 
and measure randomly in one of two complementary bases chosen by a random 
number generator orienting a half-wave plate in front of a polarizing beam 
splitter via step motors.
For the encoding, the detection of a photon in the transmitted (reflected) 
output mode of the polarizing beam splitter is set to a bit value of $0$ ($1$).
The four photons were detected by eight silicon avalanche photodiodes with 
detection efficiencies of about $40\%$. For the registration of all $16$ 
relevant four-photon coincidences, we used an eight-channel multi-photon 
coincidence unit~\cite{cc}. We observed a four-photon state rate of about 
$0.4$ per second. To build up a shared key, the acquisition time for each 
randomly chosen analysis setting was set to $1$~s. If more than one 
four-photon coincidence event was registered during that time, only the first 
one was chosen.

\begin{figure}[h!]
\begin{center}
\epsfig{file=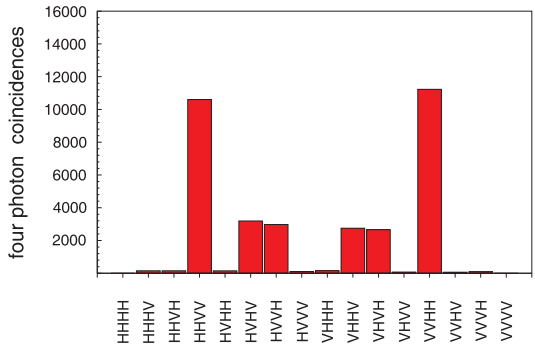, width=7.0cm}
\caption{\label{psi4hv}
Analysis of the four-photon source, where Alice, Bob Claire and David have 
chosen the $\{H,V\}$ basis for their polarization analysis. Shown are all 
possible four-photon coincidence detection events under the condition that in
each of the four output modes a photon was detected.
The four-photon correlation is $E=0.945\pm 0.002$.}
\end{center}    
\end{figure}

An analysis of the four-photon source used for the experimental 
implementation, where Alice, Bob, Claire and David have measured in the 
$\{H,V\}$ basis is shown in Fig.~\ref{psi4hv}. The data acquisition time was 
$24$ hours. Due to differences in the efficiencies of the detectors, the 
presented four-photon coincidences have been corrected without changing the 
overall raw detection rate. The corresponding four-photon correlation is 
$E=0.945\pm 0.002$, which demonstrates the high quality and stability of 
this four-photon source.

\begin{figure}[h!]
\begin{center}
\epsfig{file=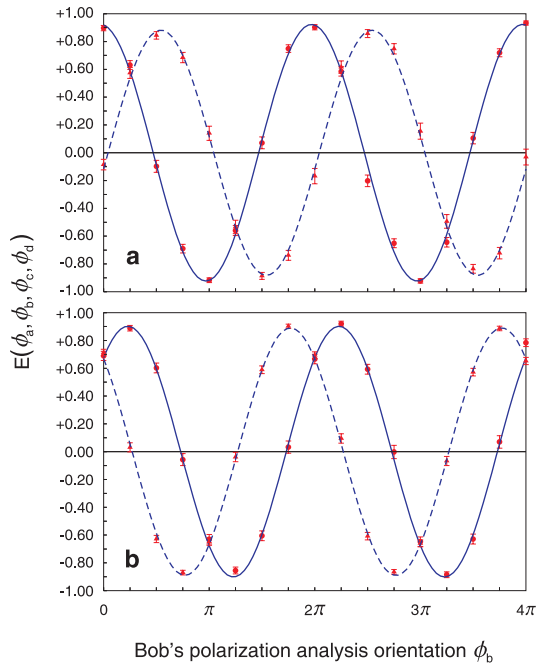, width=7.0cm}
\caption{\label{cor4} 
Four-photon polarization correlations: Alice, Claire and David have set their 
polarization analysis orientation $(\phi_{a},\phi_{c},\phi_{d})$ in $(a)$ 
either to $0$ (rounds) or $\pi/2$ (triangles) corresponding to the 
$\{\{H,V\},\{P,M\}\}$ basis set, or in $(b)$ either to $+\pi/4$ (rounds) or 
$-\pi/4$ (triangles) corresponding to the 
$\{\{22.5^{\circ},112.5^{\circ}\},\{-22.5^{\circ},67.5^{\circ}\}\}$ basis set,
while Bob varied his polarization analyzer setting $(\phi_{b})$ from $0$ to 
$4\pi$. 
The solid and the dashed lines represent numerical fits, leading in $(a)$ to 
a visibility of $V_{H/V}=92.3\% \pm 0.8\%$ and $V_{P/M}=88.2\% \pm 1.2\%$, 
and in $(b)$ to a visibility of 
$V_{22.5^{\circ}/112.5^{\circ}}=90.2\% \pm 1.1\%$ and 
$V_{-22.5^{\circ}/67.5^{\circ}}=89.0\% \pm 0.7\%$.}    
\end{center}
\end{figure}

To demonstrate the perfect four-party correlations, necessary for four-party 
QSS, we analyzed the four-photon correlations under different polarization 
analyzer orientations. In Fig.~\ref{cor4} $(a)$ the results of two measurements
are shown, where Alice, Claire and David have analyzed their photons in the 
$\{H,V\}$ $(\{P,M\})$ basis corresponding to $\phi_{a,c,d}= 0~(\pi/2)$, while 
Bob varied his analysis direction $\phi_{b}$ from $0$ to $4\pi$. The maximal 
absolute value of $E(\phi_{a},\phi_{b},\phi_{c},\phi_{d})$ can be expressed by
a visibility $V$ according to $E=V\overline{E}$, where $\overline{E}$ denotes 
the theoretical value. Fitting the experimental data leads to a visibility of 
$V_{H/V}=92.3 \pm 0.8\%$ and $V_{P/M}=88.2 \pm 1.2\%$. 
Figure~\ref{cor4} $(b)$ shows the experimental results obtained from two 
measurements, where Alice, Claire and David analyzed their photons in the 
$\{22.5^{\circ},112.5^{\circ}\}$ ($\{-22.5^{\circ},67.5^{\circ}\}$) basis 
corresponding to $\phi_{a,c,d}=\pi/4~(-\pi/4)$, while Bob varied his analyzer 
setting $\phi_{b}$. The resulting visibilities are 
$V_{22.5^{\circ}/112.5^{\circ}}=90.2 \pm 1.1\%$ and 
$V_{-22.5^{\circ}/67.5^{\circ}}=89.0 \pm 0.7\%$, respectively. The overall 
data acquisition time for each four-photon correlation function was $8.5$ 
hours. The average visibility can be translated into the QBER via 
$QBER=(1-\overline{V})/2$ leading to a QBER of $4.88\pm 0.50\%$ 
and $5.22 \pm 0.45\%$ for each basis set~\cite{qkdra}.
\begin{figure*}[t!]
\begin{center}
\epsfig{file=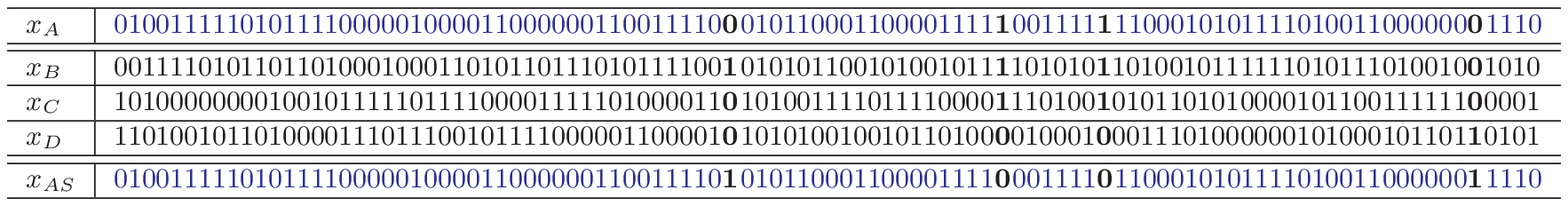}
\caption{\label{keys}
Sifted key: Shown are $100$ out of $2000$ experimentally exchanged four-party 
key bits. Bits printed in bold indicate errors. The first row shows the key of
Alice denoted by $x_{A}$. The second, third and fourth row shows each $100$ 
key bits of Bob, Claire and David denoted by $x_{B},x_{C}$ and $x_{D}$. 
As can be seen, no one of them alone and no two parties of them together are 
able to obtain the key of Alice. Only by cooperation of Bob, Claire and David 
(access set) they will be able to reconstruct the key of Alice 
via the computation of the logical XOR function 
$(x_{AS}=x_{B}\oplus x_{C} \oplus x_{D}=x_{A})$, as shown in 
the last row.}
\end{center}
\end{figure*}

For a full experimental demonstration of four-party QSS, we performed a key 
exchange according to the four-party QSS protocol described above using two 
different complementary basis sets. Using a key sifting analog to the BB84 
protocol, we have exchanged $2000$ key bits in a complete transfer time of 
about $16$ hours with a QBER of $5.20\%$. A fraction of $100$ bits of the 
sifted key is shown in Fig.~\ref{keys}. Neglecting the dead time required to 
set the polarization analysis direction and considering in addition those 
events which have been registered in the same time interval, leads to a bit 
rate of $196$ bits per hour being in well agreement with the theoretical 
expected rate. To check for eavesdropping, Alice chooses a random subset of 
$200$ bits $(10\%)$ of the sifted key and asks Bob, Claire and David for their
results at those positions. From this subset, she evaluates the QBER and 
obtains a value of $4\%$~\cite{qberf}. This value lies well below several 
known security threshold values required for two-party QKD and should therefore
be low enough to distill finally a perfect secure key~\cite{tv4qss,qkdra}.
Thus, using the remaining $1800$ bits to perform key reconciliation and privacy
amplification, Alice ends up with a totally secure key with about $200$ bits
~\cite{krp}. For the reconstruction of the final key, Bob, Claire and David 
have to combine their individual bits of the reconciled key and have to 
perform the same privacy amplification procedure as Alice.
  
Using the Bell angles ($\phi^{1,2}_{x}=\pi/4,-\pi/4$) for key sifting, while 
Bob switches every fifth measurement to $\phi^{1,2}_{b}=0,\pi/2$, we obtained
$1342$ key bits in about $21$ hours with a QBER of $5.96\%$.
Since $1/5$ of the measured data can be used to evaluate the Bell inequality 
given by Eq.~(\ref{mzbi}), we obtained a value of $S=1.78 \pm 0.07$, which is 
well above the classical limit of $1$ and very close to the theoretical value 
of $1.886$. Since the value for $S$ scales linear with the visibility, the 
achieved value of $S$ can be translated into a QBER of about $3\%$, which 
should be low enough to ensure secure quantum communication
~\cite{tv4qss,qkdra}.

We demonstrated four-party QSS via the resource of four-photon entanglement. 
We obtained bit rates of about $100$ bits per hour with QBERs of about $5\%$, 
which should be low enough to ensure perfect security. Comparing this with 
QBERs usually obtained in QKD between two parties, we obtain similar 
results demonstrating the feasibility of secure multiparty quantum 
communication via multi-photon entanglement. For future implementations, it 
would be useful to use fast optical switches for an efficient registration. 
Using in addition known techniques to increase the efficiency of the key 
exchange, this scheme can be extended to obtain the maximal reachable 
efficiency~\cite{lc}.

\begin{acknowledgments}
We would like to thank A. Sen, U. Sen and M. Zukowski for useful 
conversations. This work was supported by the DFG, 
the Bavarian high-tech initiative 
and the EU-Projects RamboQ and QAP.
\end{acknowledgments}


\begin{thebibliography}{30}
\bibitem{krr}        To remove this partial information via universal hashing, 
                     the reconciled key must be reduced by a factor of 
                     $r>2/3$.
\bibitem{tv4qss}     Until now, no unconditional security threshold value for 
                     any entanglement based QSS protocol has been strictly 
                     proven, but it seems to be reasonable to assume similar 
                     threshold values as for two-party QKD protocols.
\bibitem{bv4qss}     So far, no connection between the security of this QSS 
                     protocol and the violation of $S$, given by 
                     Eq.~(\ref{mzbi}), has been pointed out, but the value of 
                     $S$ can be translated into the QBER allowing to estimate 
                     a threshold value for $S$~\cite{tv4qss}.
\end{thebibliography}

\end{document}